\documentclass[final,5p,times,twocolumn]{elsarticle}
\usepackage{amssymb}
\usepackage{gensymb}
\usepackage{amsmath}
\usepackage{upgreek}

\begin{document}

\begin{frontmatter}

\title{Magnetic order in Co[Cr(CN)$_6$]$_{{2/3}}\cdot$$z$H$_2$O and its interplay with correlated vacancy order}

\author[ox,mat]{Rebecca E. Witte}
\author[ox]{Elodie A. Harbourne}
\author[ox,mat]{John Cattermull}
\author[ox]{Tristan N. Dolling}

\author[ox]{Simon J. Cassidy}

\author[is]{Fabio Orlandi}
\author[is]{Pascal Manuel}

\author[ox]{Andrew L. Goodwin}
\address[ox]{Department of Chemistry, University of Oxford, Inorganic Chemistry Laboratory, South Parks Road, Oxford, OX1 3QR, U.K.}
\address[mat]{Department of Materials, University of Oxford, Parks Road, Oxford, OX1 3PH, U.K.}
\address[is]{ISIS Neutron and Muon Source, Rutherford Appleton Laboratory, Harwell Campus, Didcot, OX11 0QX, U.K}

\begin{abstract}
Prussian blue analogues (PBAs) are versatile hybrid framework materials that often exhibit local and long-range departures from their cubic parent structure, influencing their functionality. Long-range magnetic order is common in PBAs, owing to magnetic interactions between transition metal ions via bridging cyanide ligands in the structure. Here we report the magnetic structure of cobalt (II) hexacyanochromate, Co[Cr(CN)$_6$]$_{{2/3}}$.$z$H$_2$O, determined using neutron powder diffraction. No evidence for magnetostriction at the magnetic ordering temperature is observed in contrast to the reported behaviour of a related PBA. Following experimental characterisation of the magnetic order of this PBA, we employ a simple Monte Carlo model to gain a better understanding of the relationship between hexacyanometallate vacancy distributions and magnetic ordering behaviour in these materials. This model shows that magnetic ordering is robust to the extent and nature of vacancy correlations, with only minor effects noticeable in the critical region, suggesting that PBAs with the same composition but different vacancy correlations might be expected to show similar magnetic behaviour. 
\end{abstract}

\begin{keyword}
magnetism  \sep Prussian blue analogues \sep short-range order \sep Monte Carlo \sep diffuse scattering \sep cobalt hexacyanochromate
\end{keyword}

\end{frontmatter}

\section{Introduction}

Prussian blue analogues (PBAs) are a diverse family of solids with a general formula of A$_{x}$P[R(CN)$_{6}$]$_{y}\cdot$ $z$H$_{2}$O (simplified to P[R]$_y$ for A-free stoichiometries) \cite{Bostrom2021}. Compositional and structural variety provide functional versatility, supporting a range of potential applications; these include uses as cathode materials \cite{Zampardi2020, Hurlbutt2018}, solid state refrigerants \cite{Ohkoshi2023}, catalysts \cite{Marquez2018}, and gas storage media \cite{Kaye2005, Thallapally2010}. 

\begin{figure}
\centering
  \includegraphics{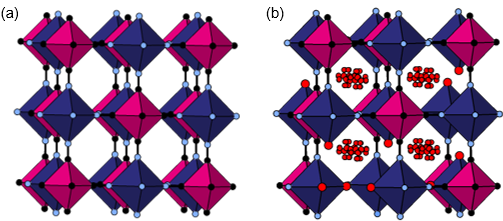}
  \caption{Representations of the structure of PBAs. (a) The structure of a defect-free Prussian blue analogue with formula P[R(CN)$_{6}$], omitting zeolitic water. The blue and pink octahedra represent P and R cations respectively. The black atoms represent carbon and the blue atoms represent nitrogen. (b) A representative structure for the P$^{\text{II}}$[R$^{\text{III}}$(CN)$_{6}$]$_{2/3}\cdot$ $z$H$_{2}$O stoichiometry. Oxygen atoms from coordinated and zeolitic water are shown in red. Zeolitic water molecules are shown at the average crystallographic positions. In a real sample, these sites are not fully occupied.}
  \label{fig1}
\end{figure}

The parent structure of PBAs is shown in Fig.~\ref{fig1}(a). P and R transition-metal sites alternate on a cubic net and are connected by cyanide ions such that their local coordination is octahedral \cite{Ludi1970}. These bridging cyanide ligands mediate superexchange coupling between neighbouring transition metal ions, enabling long-range magnetic ordering. A common PBA stoichiometry, the focus of this article, is P$^{\text{II}}$[R$^{\text{III}}$(CN)$_{6}$]$_{2/3}\cdot$ $z$H$_{2}$O (Fig.~\ref{fig1}(b)). For charge balancing, one-third of the [R$^{\text{III}}$(CN)$_{6}$]$^{3-}$ sites are vacant \cite{Cattermull2021}. Fig.~\ref{fig1}(b) shows the locations of water molecules in a PBA with P[R(CN)$_{6}$]$_{2/3}$ stoichiometry. Water is both coordinated to the P cation (in place of cyanide at hexacyanometallate vacancies) and also occupies interstitial sites within the pore network \cite{Herren1980}. Hydrated PBAs can support proton conductivity, where H-bonded networks of guest water molecules promote fast proton transport via the Grotthuss mechanism through the structure \cite{Kim2009, Yin2022, Ono2017}, aiding application as additives in proton exchange membranes \cite{Yin2022} and proton sensors \cite{Hu2022}.

There has been long-standing interest in the magnetic properties of PBAs. Magnetic characterisation of the ferromagnetic parent compound Prussian blue (Fe$^{\text{III}}$[Fe$^{\text{II}}$(CN)$_{6}$]$_{3/4}\cdot$ $z$H$_{2}$O) began as early as the 1920s \cite{Davidson1928}. Superexchange coupling interactions through bridging cyanide ligands are very weak due to the through-bond distance of over 10\,\AA{} between magnetic Fe$^{3+}$, leading to a Curie temperature of 5.6\,K \cite{Verdaguer2004}. Shorter distances of $\sim$5\,\AA{} between magnetic ions in PBAs can give rise to higher magnetic ordering temperatures with many PBAs displaying magnetic transitions between 10--150\,K \cite{Verdaguer2004}. Room temperature magnetism has even been observed in ferrimagnetic V[Cr(CN)$_{6}$]$_{0.86}\cdot$2.8H$_{2}$O \cite{Ferlay1995}. Magnetic properties of PBAs are commonly characterised by magnetometry \cite{Ohkoshi2004, Nakotte2016, Zentkova2006}, but despite interest in the magnetic properties of these materials, there are few reported direct measurements of long-range magnetic ordering from neutron diffraction.The few key examples include: determination of the magnetic unit cells of ferromagnetic Ni[Cr]$_{2/3}$ \cite{Pajerowski2015} and ferrimagnetic Mn[Cr]$_{2/3}$ \cite{Mihalik2010}, the study of the photomagnetic PBA K$_{0.27}$Co[Fe(CN)$_6$]$_{0.73}$[D$_2$O$_6$]$_{0.27}\cdot$1.42D$_2$O \cite{Pajerowski2012}, and characterisation of Cu$_{0.49}$Mn$_{0.51}$[Fe(CN)$_{6}$]$_{2/3}$, a mixed ferroferrimagnet, which displays negative magnetisation \cite{Kumar2008}. 

Whilst Co[Cr]$_{2/3}$, a ferromagnetic PBA with a Curie temperature of 28\,K at high hydration, has previously been characterised at low temperature using neutron diffraction, no magnetic Bragg peaks were observed \cite{Bhatt2013}. This is unexpected: this material has been widely characterised using magnetometry due to its interesting hydration-dependent magnetic properties where magnetic ordering has been well established \cite{Ohkoshi2004, Nakotte2016}. Understanding the relationship between structure and magnetism in Co[Cr]$_{2/3}$ would also provide comparison with related ferrimagnetic V[Cr]$_{2/3}$. The latter is reported to exhibit magnetostriction at its Curie temperature (313\,K) accompanied by an increase in the activation energy for proton conductivity, possibly due to the Grotthuss mechanism being disrupted by the change in the unit cell volume and therefore the water locations within the hydrogen bonding network \cite{Ohkoshi2010}.  

It has been suggested that the absence of magnetic Bragg intensity in the neutron diffraction patterns of Co[Cr]$_{2/3}$  below its magnetic ordering temperature is related to diffuse scattering from disordered water and vacancies \cite{Bhatt2013}. PBAs with a hexacyanometallate vacancy fraction of one-third cannot support complete isolation of vacancies, leading to the formation of vacancy neighbour pairs that may connect to form wider pore networks throughout the structure \cite{Simonov2020}. These vacancy distributions are known to be non-random in single crystals of PBAs, where different vacancy correlations relate to changes in composition and synthesis conditions \cite{Simonov2020}. The diffuse scattering, characteristic of these correlations, is mostly hidden by the orientational averaging in powder diffraction, although it can appear as broad maxima centred at primitive superlattice positions \cite{Simonov2020, Grandjean2016}. One such broad feature was observed for Cu$_{0.49}$Mn$_{0.51}$[Fe(CN)$_{6}$]$_{2/3}$, but magnetic Bragg peaks were still visible at low temperature \cite{Kumar2008}. Hence there is no reason to expect vacancy order necessarily precludes observation of magnetic Bragg reflections. 

The influence of the distribution of vacancies in PBAs on collective properties such as mass transport is known \cite{Simonov2020}, but to our knowledge, the relationship between vacancy correlations and magnetism has not been explored. The connectivity between vacancies in pore networks contributes to the conductance \cite{Simonov2020}, so we might expect changes in local superexchange pathways between magnetic ions to influence long-range magnetic order. Understanding whether magnetic order depends upon vacancy order will aid the analysis of the magnetic behaviour of PBAs synthesised following different methods, which could have different vacancy correlations.   

In this study, we show that the emergence of bulk ferromagnetism in Co[Cr]$_{2/3}$ is indeed accompanied by magnetic Bragg scattering, which we use to determine the ground state magnetic structure. High resolution variable temperature neutron scattering measurements across the magnetic transition suggest that there is no significant variation in unit cell dimensions, suggesting that this material does not demonstrate magnetostriction. We also introduce a simple Monte Carlo (MC) model to investigate the importance of random versus correlated vacancy distributions on magnetic ordering behaviour in PBAs; our analysis suggests that any changes to  magnetic ordering are minimal, although small deviations are present close to the transition temperature. 

Our paper is arranged as follows. After an outline of the methods used in this study, we show our structural and magnetic characterisation of Co[Cr]$_{2/3}$. We proceed to discuss the implementation of the MC model to investigate the effect of random versus correlated vacancy distributions on the magnetic ordering behaviour of PBAs. Our paper concludes that bulk ferromagnetism is observable via neutron diffraction for Co[Cr]$_{2/3}$ and any vacancy correlations present do not impact the observed magnetic behaviour. 

\section{Methods}

\subsection{Cobalt (II) hexacyanochromate}

\label{synth}

The synthesis method used followed closely those previously reported for Co[Cr]$_{2/3}$ \cite{Sato2003, Ohkoshi2004, JTBostrom2019}. Heating of the sample was avoided during synthesis because this can induce linkage isomerisation \cite{Reguera2010, Brown1969}. Stoichiometric solutions of K$_{3}$Cr(CN)$_{6}$  and CoSO$_{4}\cdot$$_{}7$H$_{2}$O dissolved in distilled water were prepared. The cobalt solution was added to the K$_{3}$Cr(CN)$_{6}$ solution dropwise over 15\,min whilst stirring at room temperature. The pink mixture was left to stir at room temperature for 1\,h before standing for 1\,h. The peach precipitate was separated by centrifugation, followed by three washes with distilled water. The peach precipitate was left to air dry at ambient temperature and pressure over several days. This procedure was followed to synthesise five batches of Co[Cr]$_{2/3}$ (labelled sample A--E).

\subsection{Magnetometry}

A superconducting quantum interference device (SQUID) magnetometer was used to measure the magnetisation of sample A at different temperatures. Zero field cooled (ZFC) and field cooled (FC) magnetisation data were collected for sample A between $2\,$K and $300\,$K. 

\subsection{Powder X-ray diffraction (PXRD)}

For initial phase identification, PXRD measurements were carried out using a Bruker D8 Advance Eco diffractometer, with a Cu source with X-ray wavelength of $1.5406$\,\AA{}. The diffraction patterns were recorded over an angular range $2\theta=10$--$80\,\degree$ with sample spinning using a 20 minute scan. 

A Pawley refinement approach, using the TOPAS-Academic v6 software package \cite{TOPAS}, was employed, using the modified Thompson-Cox-Hasting pseudo-Voigt (TCHZ) peak shape function \cite{Young1993}.

\subsection{Neutron powder diffraction (NPD)}

NPD measurements were carried out at the WISH beamline at ISIS. The combined Co[Cr]$_{2/3}$ sample (containing samples A--E) (0.8708\,g) was loaded into a thin-walled 6\,mm vanadium can. This sample was still hydrogenous rather than deuterated due to concerns about the PBA rehydrating with atmospheric moisture after removal of the sample from the vacuum oven, potentially giving a mixture of H and D on interstitial water molecules. The upper section of the can was wrapped in a cadmium mask. The sample was placed into an Oxford Cryostat (WISH 30\,L Variox) at 100\,K. The sample volume was evacuated before filling with helium exchange gas. This process was repeated a second time before reducing the pressure to 100\,mbar. 

The sample was first cooled from 100\,K to 1.5\,K. Data were collected following a schedule: longer measurements ($\sim250$\,mins) at 1.5\,K and 40\,K; then shorter measurements ($\sim15-40$\,mins) on warming at 10\,K, 15\,K, 17.5\,K, 20\,K, 22\,K, 23\,K, 24\,K, 25\,K, 26\,K, 27\,K 28\,K, 30\,K, and 35\,K. 

Refinements employing the TOPAS-Academic v6 software package \cite{TOPAS} were completed simultaneously over five data sets, corresponding to different detector banks present at the WISH beamline \cite{chapon_wish_2011}. The lattice parameter was fixed across the data banks, but the scale and peak shapes were allowed to refine individually in each bank using a time-of-flight peak shape function. The magnetic structure of the PBA was refined using the difference between the raw intensities at 40\,K and 1.5\,K, selected to be above and below the Curie temperature. Modes consisting of linear combinations of magnetic moments from Co or Cr atoms (from irreducible representations of the space-group symmetry of a crystal) were obtained using ISODISTORT \cite{ISODISTORT1, ISODISTORT2}. The scale of each bank was fixed using values obtained from a constrained Rietveld refinement constraining C--N bond lengths and the vacancy fraction to allow estimation of the magnetic moment magnitudes. 

\subsection{Monte Carlo (MC) simulations}
\label{MC}

\begin{figure}
\centering
  \includegraphics{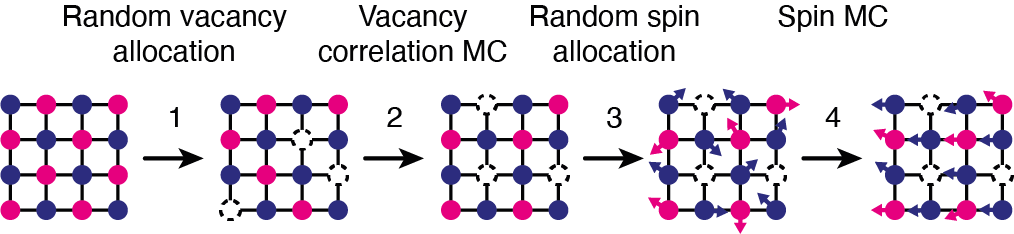}
  \caption{Summary of MC modelling approach for configurations with correlated vacancies. P sites are shown as blue circles, present R sites are shown as pink circles, R-site vacancies are shown as dotted white circles, and C and N atoms are omitted for clarity. (1) Vacancies are randomly allocated to R sites in accordance with the chosen vacancy fraction. (2) The vacancy correlation MC changes the vacancy distribution in accordance with the chosen Hamiltonian \cite{Simonov2020}. (3) Heisenberg spins are allocated to the magnetic sites with random initial orientations. (4) The magnetic MC follows a simulated annealing algorithm which models ferromagnetic ordering.}
  \label{fig2}
\end{figure}

We used a Metropolis MC approach to explore the interplay of vacancy and magnetic order \cite{Metropolis1949}. Two successive MC models were implemented: first, vacancy configurations were generated; then, spins were assigned to the occupied lattice positions before employing a spin MC. The overall process for considering configurations with vacancy correlations in the magnetic MC is summarised in Fig.~\ref{fig2}. 

\subsubsection{Generating vacancy configurations}

Configurations containing correlated vacancies were generated using custom MC code employing the Hamiltonian from the model for vacancy distributions from Ref.~\citenum{Simonov2020}. In that study, effective vacancy interactions were governed by two interactions $J_1$ and $J_2$. Here we recast these into derived parameters $A = \frac{J_{1}}{J_{2}}$ and $B = \frac{T}{J_{2}}$, where $T$ is the Monte Carlo temperature. Values of $A$ and $B$ were chosen for consistency with Ref.~\citenum{Simonov2020}, enabling direct comparison of single-crystal diffuse scattering patterns calculated from configurations resulting from the MC model with experimental data. Starting with an initial configuration representing the PBA structure, including P and R sites each arranged on their respective face-centred cubic (FCC) sublattice, one-third of randomly selected R site positions were labelled as vacancies. Configurations included $9 \times 9 \times 9$ unit cells, containing a total of 2916 P sites. The initial energy of the configuration was calculated using a predetermined $A$ value. Periodic boundary conditions were applied to determine the nearest neighbours of sites \cite{Asselin2010}.

Each MC move consists of the swapping of one occupied and one vacant R site chosen at random before calculating the associated energy change. The final configuration was recorded after a total of 360 $\times$ number of P sites moves, which was determined to be sufficient to reach equilibrium. 30 independent vacancy configurations were generated for each pair of $A$ and $B$ values. 

\subsubsection{Magnetic MC model}

\label{spinMC}

The code used to implement the MC algorithm was adapted from Ref.~\citenum{paddison_ewald}. Changes were required to allow for the introduction of vacancies into a supercell to enable investigation of vacancy ordering.

The magnetic MC simulation used a configuration with either a random or correlated vacancy distribution. R sites were removed from the configuration to achieve the desired vacancy fraction, $x_\textrm{vac}$.  Each occupied P and R site was treated as a magnetic site and decorated with a classical Heisenberg spin in a random initial orientation. All spins were approximated to share the same magnitude as this applies for the spin-only magnetic moments in Co[Cr]$_{2/3}$, where S = $\frac{3}{2}$ for both Co$^{\text{II}}$ and Cr$^{\textrm{III}}$.

The magnetic interactions in PBAs are considered using the Heisenberg Hamiltonian
\begin{equation}
H_{i} = -\frac{1}{2}\sum_{ij}J\textbf{S}_{i}\cdot\textbf{S}_{j} ,
\label{MagHamiltonian}
\end{equation}
where $J$ is the strength of the superexchange interaction between a metal site $i$ and its first nearest neighbour $j$, and \textbf{S} is the spin associated with a transition metal site \cite{dasilva_mean-field_2012}. Second nearest neighbours and beyond are excluded from these simulations due to the greater through-bond distance between the magnetic sites \cite{Verdaguer2004}. Simulations were carried out at effective temperatures $T_{\textrm{MC}} = T/J$, using $J>0$ to model ferromagnetic interactions. 

We used a simulated annealing approach, monitoring the evolution of long-range magnetic order via the order parameter, $\upphi$, given by
\begin{equation}
\upphi = \frac{1}{N_{s}}\bigg|\sum_{i=1}^{N_{s}}\textbf{S}_{i}\bigg|,
\label{MagHamiltonian2}
\end{equation}
where $N_{s}$ is the total number of spins in the configuration. 

\section{Results and discussion}

\subsection{Cobalt (II) hexacyanochromate crystal structure}

\begin{figure}
\centering
  \includegraphics{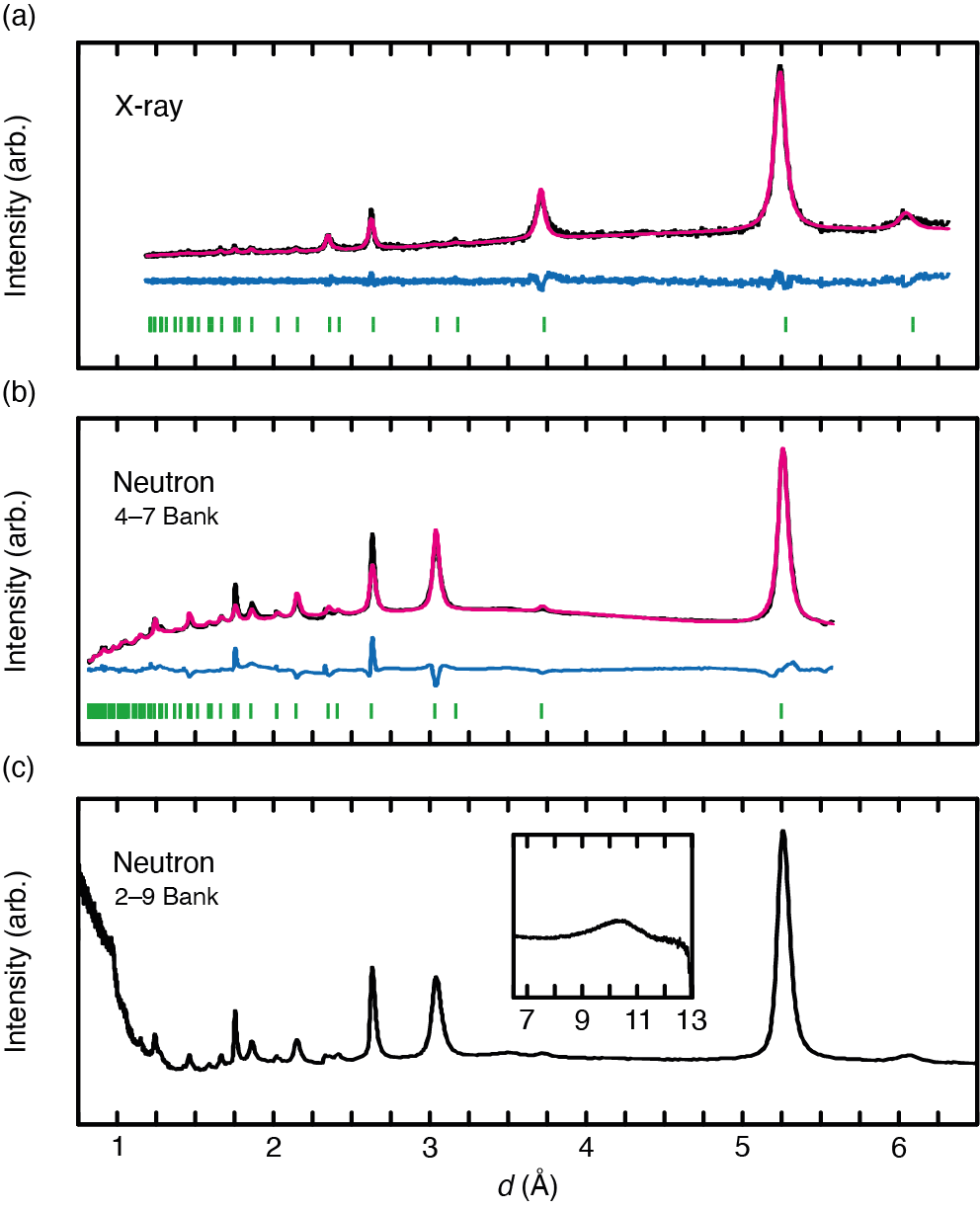}
  \caption{Co[Cr]$_{2/3}$ diffraction patterns. (a) PXRD data at ambient temperature and corresponding Pawley fit for sample A. (b) NPD data and corresponding Rietveld fit collected at 40\,K for bank with average 2$\theta$ of 121.6\,$^\circ$ at the WISH beamline, ISIS, using combined sample A--E. (c) NPD data collected at 40\,K for for bank with average 2$\theta$ of 58.3\,$^\circ$ at the WISH beamline, ISIS, for combined sample A--E. The inset highlights the (100) position. Experimental data shown in black, fit in pink, the differences curve in blue and tick marks in green.}
  \label{fig3}
\end{figure}

Cobalt hexacyanochromate was synthesised following a simple coprecipitation procedure (section~\ref{synth}) in multiple batches, allowing sufficient quantities of sample to be obtained for the neutron diffraction measurements. The peach powder collected matches the colour expected from the literature \cite{Sato2003}. A representative PXRD pattern obtained is shown in Fig.~\ref{fig3}(a). The unit cell parameters resulting from Pawley refinement in the $Fm\bar3m$ space group of the collected PXRD data for each batch of Co[Cr]$_{2/3}$ varied only slightly between 10.545$\,$\AA\, and 10.560$\,$\AA, which is in accord with the values expected from a sample with high hydration \cite{Ohkoshi2004}. All peaks in the PXRD pattern are accounted for by the $Fm\bar3m$ model, suggesting the absence of any significant quantities of impurity phases. 

At the WISH beamline, NPD data were obtained using a combined sample of all five Co[Cr]$_{2/3}$ batches. An example neutron diffraction pattern recorded at 40\,K is displayed in Fig.~\ref{fig3}(b). At 40\,K, the sample is above the Curie temperature, meaning that the Bragg peaks in the diffraction pattern arise from nuclear scattering. The refined Rietveld parameters are displayed in Table~\ref{Riet}. The data did not allow reliable refinement of O/H positions and occupancies from interstitial water so these were excluded from the model. The difficulty of obtaining stable Rietveld refinements for PBA models even with high resolution diffraction measurements is well known in the field, and arises from the very high crystal symmetry \cite{Nielsen_2026}. The hexacyanometallate vacancy fraction was fixed to one-third, with each vacancy refined with a coordinated O atom from the coordinated water. The C --N bond length was constrained to values between $1.10-1.18\,$\AA{} and the thermal parameters of each atoms were constrained to be equal. The overall fit accounts for all peaks present in the diffraction pattern, with deviations in intensity originating from the approximations made. 
\begin{table}
\small
    \caption{Crystallographic parameters for the $Fm\bar3m$ structure of Co[Cr(CN)$_6$]$_{2/3}\cdot$$z$H$_2$O at 40\,K determined using neutron diffraction. The constrained Rietveld fit fixed the hexacyanochromate vacancy fraction to one-third and omitted interstitial water. The experimental data was collected at WISH beamline, ISIS.}
    \label{Riet}
    
        \begin{tabular}{l|lllll}
            \hline
            $a$/\AA&\multicolumn{5}{l}{10.4922(4)}\\
            $R_{wp}$&2.05\\\hline
            Atom&$x$&$y$&$z$&occ&$B_{\rm eq}$/\AA$^2$\\\hline
            Co&0.5&0.0&0.0&1&1.43(8)\\
            Cr&0.0&0.0&0.0&0.6667&1.43(8)\\
            C&0.1896(5)&0.0&0.0&0.6667&1.43(8)\\
            N&0.3024(14)&0.0&0.0&0.6667&1.43(8)\\
            O1&0.315(5)&0.0&0.0&0.3333&1.43(8) \\
            \hline
        \end{tabular}
    
\end{table}

We note that a broad diffuse feature was also observed near the (forbidden) (100) reflection position (shown in Fig.~\ref{fig3}(c)), which has also been observed in previous NPD studies of PBAs \cite{Bhatt2013, Kumar2008, bhatt_multifunctional_2025}. Features at the (100) position have also been reported for disordered rocksalt cathode materials, which can display layered-like local cation ordering despite the average $Fm\bar3m$ space group \cite{wang_exploring_2026}. 

\subsection{Magnetic behaviour}
\label{mag_section}

\subsubsection{Magnetometry}

\begin{figure}
\centering
  \includegraphics{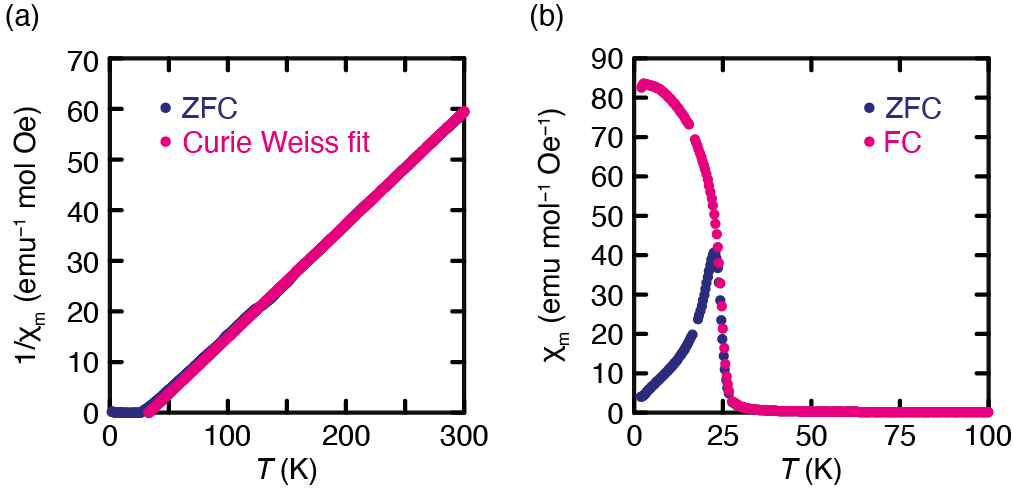}
  \caption{SQUID data for sample A using an applied magnetic field of 100\,Oe. (a) Inverse molar susceptibility 1/$\chi_{\textrm{m}}$ against temperature for ZFC data compared with Curie Weiss fit. (b) Molar susceptibility per formula unit $\chi_{\textrm{m}}$ against temperature. All analysis assumes high hydration.}
  \label{fig4}
\end{figure}

The magnetic properties of a sample of batch A cobalt hexacyanochromate were determined using SQUID magnetometry. The magnetometry measurements are summarised in Fig.~\ref{fig4}. At high temperature, the inverse molar susceptibility has a linear relationship with temperature, suggesting Curie--Weiss behaviour. The Curie constant, $C$, and Curie--Weiss temperature, $\Theta_{\textrm{CW}}$,  derived from a linear fit to the inverse molar susceptibilities at high hydration (Co$^{\text{II}}$[Cr$^{\text{III}}$(CN)$_{6}$]$_{2/3}\cdot$ 5H$_{2}$O---chosen to reflect reported hydration values for the refined lattice parameter and in accord with peach appearance \cite{Ohkoshi2004}) are summarised in Table~\ref{CurieT}. Only data collected above $250$\,K (around 10 times the expected range of Curie temperature values) were used to derive the Curie--Weiss temperature to strike a balance between using sufficient data points to reduce error and only including data from the paramagnetic region.

In Fig.~\ref{fig4}(a), the Curie--Weiss law is shown to closely fit the inverse molar susceptibility ZFC data. The positive value of $\Theta_{\textrm{CW}}$ implies the presence of dominant ferromagnetic interactions \cite{Blundell2001}. The Curie temperature was recorded as the temperature at which a discontinuity is observed in the gradient of the molar susceptibility. In unfrustrated ferromagnetic materials, $\Theta_{\textrm{CW}}$ is often approximately equal to the Curie temperature \cite{Mugiraneza2022}. This is the case here: the Curie temperature, $T_{\textrm{C}}$, was determined to be $\sim$28\,K from the ZFC and FC data assuming high hydration, which is close to the $\Theta_{\textrm{CW}}$ values of 32.8(2)\,K and 30.8(7)\,K using the ZFC and FC data respectively. The difference between these values can be quantified using the frustration parameter, $f$, which can be found by
\begin{equation}
f = \frac{|\Theta_{\textrm{CW}}|}{T_{\textrm{C}}},
\label{Frustration}
\end{equation}
where large $f$ values suggest a material is frustrated \cite{Ramirez1991}. For the ZFC and FC data, $f$ is close to 1, indicating that Co[Cr]$_{2/3}$ is not frustrated. The recorded Curie temperature is close to the reported $T_{\textrm{C}}$ for cobalt hexacyanochromate at high hydration (28\,K), and falls within the range of reported $T_{\textrm{C}}$ values that include low hydration (22\,K), providing supporting evidence that the PBA synthesised is the correct compound \cite{Ohkoshi2004}. 

\begin{table}
\small
  \caption{Values determined from SQUID measurements assuming high hydration.}
  \label{CurieT}
  \begin{tabular*}{0.48\textwidth}{@{\extracolsep{\fill}}lll}
    \hline
     Cooling method & $\Theta_{\textrm{CW}}$ (K)&$C$ (emu\,K\,mol$^{-1}$)  \\
    \hline
    ZFC&32.8(2)&4.487(4)\\
    FC&30.8(7)& 4.656(13)\\

    \hline
  \end{tabular*}
\end{table}

\subsubsection{Magnetostriction}

\begin{figure*}
\centering
  \includegraphics[width=\textwidth]{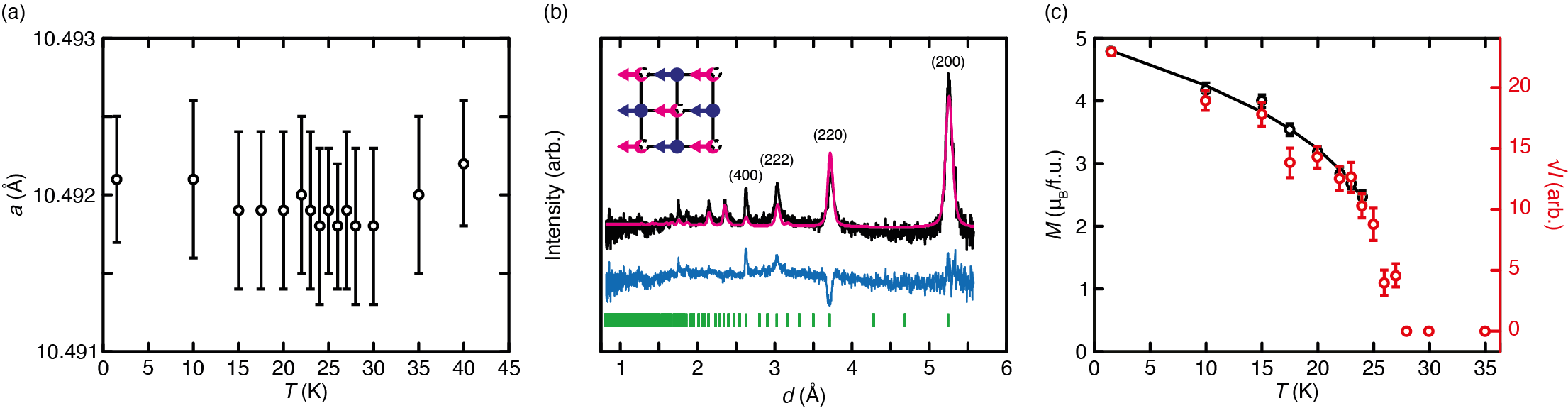}
  \caption{Magnetism results derived from neutron scattering for Co[Cr]$_{2/3}$. (a) Variation of unit cell parameter, $a$, as a function of temperature. (b) Temperature-subtracted neutron diffraction data showing magnetic Bragg reflections, with a fit corresponding to a magnetic unit cell where Co and Cr moments are parallel. The choice of the displayed moment direction relative to the atomic positions is arbitrary. Subtracted NPD data ($1.5\,\textrm{K}-40\,\textrm{K}$) for bank with average 2$\theta$ of 121.6\,$^\circ$ collected at the WISH beamline, ISIS, is displayed in black. The fit to the data is shown in pink, the difference curve is shown in blue, and the tick marks are green. Co and Cr atoms are shown in blue and pink respectively. All other atoms are omitted from the refinement. Magnetic moments are indicated with an arrow. (c) Variation of total magnetic moment as a function of temperature with fit of $M = \alpha(1-\frac{T}{T_{\textrm{C}}})^{\beta}$ to the data (open black circles). Superimposed as open red symbols are the integrated intensities of the $(200)$ reflection.}
  \label{fig5}
\end{figure*}

The unit cell parameter values obtained by Rietveld refinement of the NPD patterns collected across a temperature range from 1.5\,K to 40\,K are shown in Fig.~\ref{fig5}(a). 
No meaningful variations in the unit cell parameter are observed with increasing temperature. $Fm\bar3m$ was an appropriate space group at all recorded temperatures, indicating that there are no anisotropic changes to the unit cell. Overall, this neutron experiment provides no evidence of magnetostriction which is in contrast to the study of V[Cr]$_{2/3}$ \cite{Ohkoshi2010}. We note that the degree of magnetostriction reported in \cite{Ohkoshi2010} is two orders of magnitude greater than the resolution-limited uncertainty in our own measurements. In this experiment, the evidence does not support the mechanism of magnetostriction influencing proton conductivity in Co[Cr]$_{2/3}$ as was proposed for V[Cr]$_{2/3}$ \cite{Ohkoshi2010}, suggesting the `spin-ionic' coupling behaviour may not be applicable to the wider hexacyanochromate PBA family.

\subsubsection{Magnetic unit cell}

To probe the signature (if any) of long-range magnetic order, we compared NPD data collected above and below $T_{\textrm{C}}$. The temperature-subtracted data ($1.5\,\textrm{K}-40\,\textrm{K}$) show increased Bragg peak intensity at 1.5\,K compared to 40\,K, consistent with magnetic Bragg scattering (Fig.~\ref{fig5}(b)). The magnetic scattering intensity coincides with peaks associated with the $Fm\bar3m$ nuclear unit cell, with the largest increased peak intensity observed at all-even ($hkl$) reflections. The absence of additional magnetic reflections that do not align with the nuclear unit cell indicates that the magnetic and nuclear unit cells are commensurate, supporting either ferromagnetic or ferrimagnetic order. 

Irreducible representations of magnetic modes corresponding to an $Fm\bar3m$ unit cell, with Co and Cr occupancies of 1 and 2/3 respectively, were obtained using ISODISTORT \cite{ISODISTORT1, ISODISTORT2} and fitted to the subtracted NPD data. m$\Gamma_{4}^{+}$ modes concerning Co and Cr were considered, where all moments from a specified site align along a chosen direction. The best fit was obtained using a magnetic unit cell with Co and Cr moments constrained parallel to the same lattice direction (Fig.~\ref{fig5}(b)), supporting long-range ferromagnetic order. Because our data are for a powder sample, and because there is no measurable magnetostriction (and the system maintains cubic symmetry), we do not have any direct sensitivity to the magnetisation direction; nor can we assign the relevant magnetic space group. Nevertheless the ferromagnetic unit cell was subsequently used to refine the Co and Cr magnetic moments across the temperature range of the experiment, using data obtained by subtracting the 40\,K pattern from the lower-temperature patterns. 

The change in the total magnetic moment per formula unit of Co[Cr(CN)$_{6}$]$_{2/3}$ with temperature is shown in Fig.~\ref{fig5}(c). To obtain an estimate of $T_{\textrm{C}}$, the data was fit to the relationship $M = \alpha(1-\frac{T}{T_{\textrm{C}}})^{\beta}$ where $\alpha$ and $\beta$ are constants and $M$ is the magnetisation \cite{Binney1992}. The $T_{\textrm{C}}$ value obtained from this approach was 26(1)\,K, which is in alignment with the $T_{\textrm{C}}$ recorded via SQUID magnetometry. The values of $\mu_{\textrm{Co}}$ and $\mu_{\textrm{Cr}}$ were refined to 2.70(3)\,$\mu_{\textrm{B}}$ and 3.13(4)\,$\mu_{\textrm{B}}$ respectively at 1.5\,K, which are close to the expected value of 3\,$\mu_{\textrm{B}}$ for an ion with three unpaired spins at 0\,K (neglecting possible impacts from covalency or unquenched orbital angular momentum). The total moment for Co[Cr]$_{2/3}$ of 4.79(5)\,$\mu_{\textrm{B}}$ closely matches the literature value of 4.7\,$\mu_{\textrm{B}}$ \cite{Ohkoshi2004}.

\subsection{Influence of vacancy distribution on magnetic order}

The NPD results in section~\ref{mag_section} provide evidence for the onset of long-range magnetic order at low temperature. Here we revisit the observation of the broad feature at (100) implying the presence of vacancy correlations to ask whether we might expect these to influence the magnetic behaviour of the Co[Cr]$_{2/3}$ sample (and PBAs more widely). To this end, we employed MC modelling to compare the onset of magnetic order between configurations with random vacancy distributions and those with correlated vacancies. Two successive MC simulations were used as part of this approach. First, an MC algorithm was used to generate a configuration of vacancy distributions using the Hamiltonian from Ref.~\citenum{Simonov2020}. And second, spins were assigned to non-vacant sites in a spin MC simulation carried out to explore magnetic ordering. 

\subsubsection{Spin MC: random vacancies}

\begin{figure}[t]
\centering
  \includegraphics{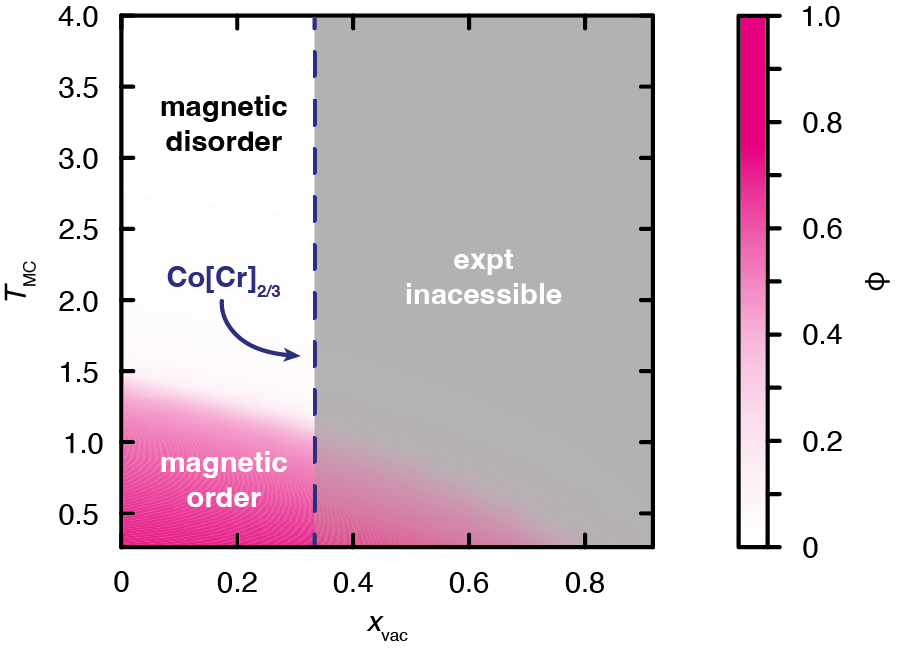}
  \caption{Magnetic MC results for input configurations with randomly distributed vacancies for a range of $x_{\textrm{vac}}$ values from 0 to 11/12. Results shown are an average of 30 independent MC runs. $\upphi$ as a function of $T_{\textrm{MC}}$ and $x_{\textrm{vac}}$.}
  \label{fig6}
\end{figure}

Initially, the spin MC simulation used input configurations containing randomly distributed vacancies. Data were collected over 30 independent MC runs for each vacancy fraction between 0 and $11/12$, in steps of $1/12$. The results are summarised in Fig.~\ref{fig6}.
On cooling, a magnetic ordering transition is observed; the transition temperature ($T_{\textrm{C}}$) is reduced for higher vacancy concentrations. This trend follows what we might expect from mean field theory, whereby the magnetic ordering temperature is directly proportional to the average coordination number of magnetic sites \cite{Verdaguer2004}. Increasing vacancy fractions leads to a linear decrease in average coordination number, so $T_{\textrm{C}}$ could also be expected to decrease linearly.

Percolation theory could explain the lack of a clear transition at the highest vacancy fractions \cite{Broadbent1957}. Onset of long-range magnetic order requires the formation of a connected pathway of magnetic sites from one side of the configuration to the other \cite{Gawron1991}. The percolation threshold is defined as the fractional occupancy of sites at which an infinite cluster forms on an infinite lattice \cite{Shante1971}. In this system, the relevant percolation threshold is that for an FCC lattice, considering both first and second nearest neighbours. This threshold is appropriate because R-site vacancies are randomly distributed on an FCC sublattice, whilst P sites are always present. To form a cluster spanning the lattice, occupied R sites must connect via the P sites. The site percolation threshold in this case is 0.136, which corresponds to a vacancy fraction of 0.864 \cite{Gawron1991}. The vacancy fraction above this threshold (0.917) does not appear to reach long-range magnetic order. 0.833 is below the maximum vacancy fraction to reach the percolation threshold, yet does not reach long-range magnetic order. The lack of consideration of percolation in mean field theory accounts for the non-linear relationship between vacancy concentration and $T_{\textrm{C}}$. However, as vacancy fractions exceeding one-third are theoretical rather than experimentally observed, the deviations from linear behaviour at the highest vacancy concentrations are not consequential for the application of this model \cite{Bostrom2024}.  

\subsubsection{Spin MC: correlated vacancies}

We generated a set of configurations with vacancy ordering deviating from the random case for PBAs following P[R]$_{2/3}$ stoichiometry using the vacancy correlation MC \cite{Simonov2020}. Configurations have assigned values of $A$ and $B$ as defined in section~\ref{MC}, where smaller values of $A$ correspond to a higher weighting of the centrosymmetric term in the Hamiltonian (as would be favoured for PBAs with Jahn--Teller active P cations) and larger values of log$B$ relate to higher effective temperatures used during the simulation.

Co[Cr]$_{2/3}$ was chosen as an initial test case to compare with the magnetic ordering data for randomly distributed vacancies. The first challenge was to determine a suitable set of values to use for parameters $A$ and log$B$, which need to be tuned to generate models with the correct vacancy distribution. To our knowledge, Co[Cr]$_{2/3}$ has no recorded single-crystal diffuse scattering data that could be compared with scattering from simulated configurations to determine appropriate $A$ and log$B$ values for generating experimentally-accurate vacancy distributions. Comparison of PBAs with a common P cation suggests that they have similar $A$ values (e.g. values between 0.75 and 1 were assigned for Mn[Fe]$_{2/3}$, Mn[Co]$_{2/3}$ and Mn[Mn]$_{2/3}$) \cite{Simonov2020}. The only previously assigned P = Co PBA was Co[Co]$_{2/3}$, where $A$ = 1 \cite{Simonov2020}. log$B$ is more challenging to determine: the same PBA synthesised under different conditions can lead to very different log$B$ values (two different samples of Mn[Mn]$_{2/3}$ were assigned log$B$ = 0.1 and log$B$ = 0.4) \cite{Simonov2020}. The $A$ and log$B$ values assigned to Co[Co]$_{2/3}$ in Ref. \citenum{Simonov2020} (1 and 0.25 respectively) were therefore assumed to reflect potential parameter values for Co[Cr]$_{2/3}$.

\begin{figure}
\centering
  \includegraphics{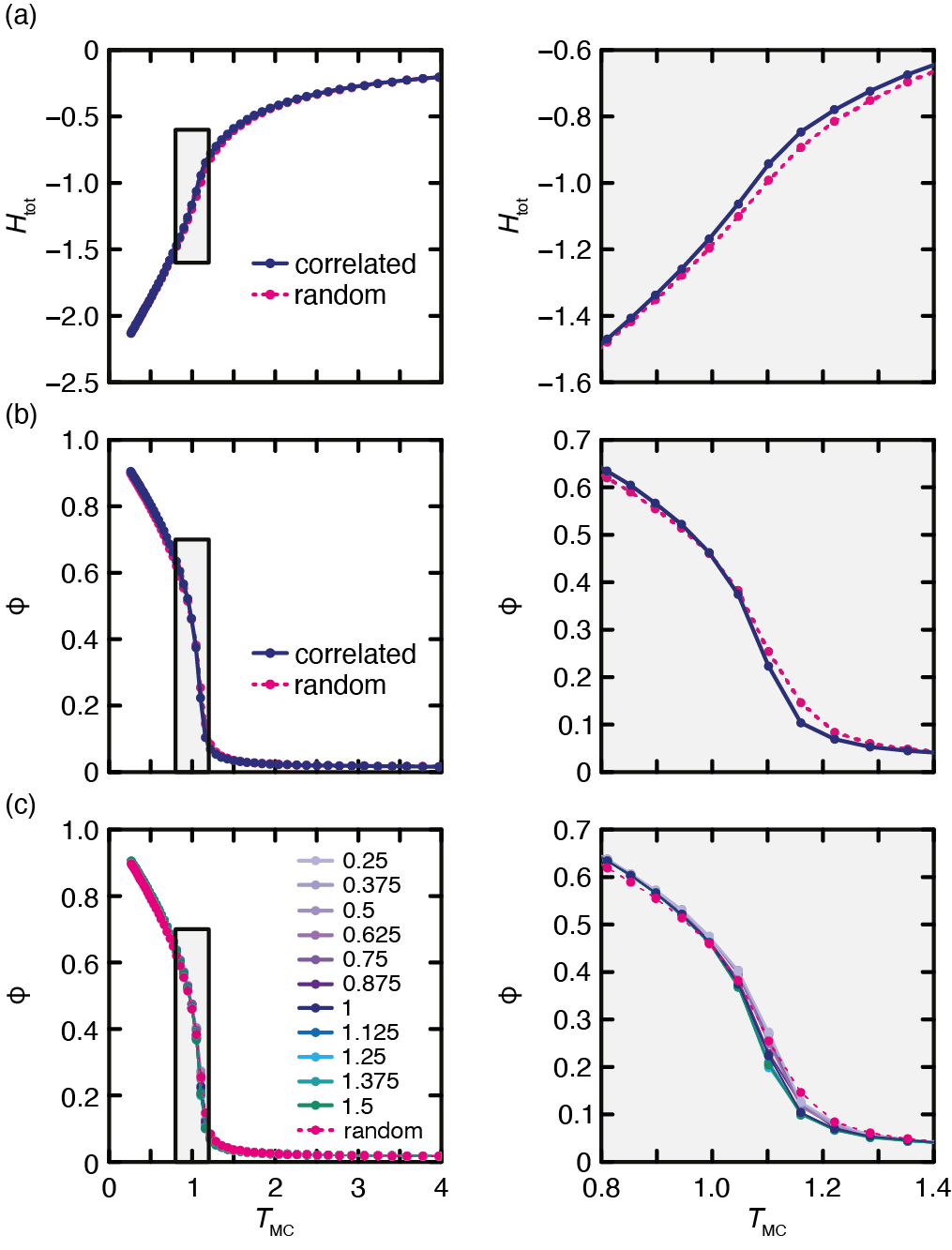}
  \caption{(a--b) Monte Carlo results for a P[R]$_{2/3}$ system with a random vacancy distribution vs Co[Cr]$_{2/3}$ with a correlated vacancy distribution : (a) Total energy per magnetic site, $H_{\textrm{tot}}$, and (b) $\upphi$ as a function of $T_{\textrm{MC}}$. (c) $\upphi$ as a function of $T_{\textrm{MC}}$ for a range of $A$ values where log$B = 0.25$ and $x_{\textrm{vac}} = 1/3$. Results shown are an average of 30 independent MC runs. Standard uncertainties are smaller than the symbols.}
  \label{fig7}
\end{figure}

The results from the magnetic MC for the random and correlated vacancies for Co[Cr]$_{2/3}$ are compared in Fig.~\ref{fig7}(a--b).  
There is essentially no difference between the random and correlated vacancies cases, suggesting that vacancy distribution does not impact magnetic order for Co[Cr]$_{2/3}$. Both before ($T_{\textrm{MC}} > 1.5$) and after ($T_{\textrm{MC}} < 0.8$) the onset of long-range magnetic order there is no significant difference between the order parameters and MC energies for the random and correlated vacancies. At the onset of magnetic order, there is a small deviation between random and correlated vacancies, with random vacancies providing a very slightly greater value of the order parameter and a more negative mean MC energy per site (suggesting spin vectors have closer alignment with their first nearest neighbours) for the same temperature.

To further explore how vacancy correlations might influence magnetic order near the transition temperature, magnetic MC simulations were carried out using parameters of log$B$ = 0.25 and $A$ from 0.25 to 1.5 in steps of 0.125. For all values of $A$ tested, the difference between the onset of magnetic order observed for random and correlated vacancies remains small, with the only notable differences seen near the transition (Fig.~\ref{fig7}(c)). There appears to be a subtle trend dependent on $A$ at the magnetic transition: small values of $A$ (corresponding with a higher preference for centrosymmetric P sites) lead to larger order parameter values near the transition than larger values of $A$. This suggests that magnetic ordering behaviour in PBAs at the transition may be impacted (if only extremely weakly) by the specific vacancy distribution and is therefore related to the PBA chemistry. The robustness of behaviour we observe is likely due to the high connectivity of superexchange pathways through the lattice that persists whether or not vacancies are present. 

\section{Concluding remarks}

The magnetic characterisation of cobalt hexacyanochromate has provided two main results. First, there is no evidence of magnetostriction in Co[Cr]$_{2/3}$, in contrast to the related PBA V[Cr]$_{2/3}$ \cite{Ohkoshi2010}. This suggests that the proposed mechanism of a change in unit cell volume at a magnetic ordering transition impacting proton conductivity \cite{Ohkoshi2010} is not applicable to Co[Cr]$_{2/3}$. Second, magnetic Bragg scattering was observed in the neutron diffraction pattern at temperatures below the Curie temperature. This measurement provides evidence of long-range magnetic order in Co[Cr]$_{2/3}$ which has been absent in previous neutron diffraction work \cite{Bhatt2013}. The refined magnetic unit cell supports the presence of long-range ferromagnetic order, matching expectations from magnetometry measurements recorded in both this study and previous work \cite{Ohkoshi2004, Sato2003, Bhatt2013, Zentkova2006}. It is possible that magnetic Bragg peaks could have been present in the work by Bhatt et al. \cite {Bhatt2013} but it is challenging to deduce without a difference plot between the high and low temperature neutron scattering data given the large background. 

Overall, the MC results suggest that local vacancy distribution has no clear effect on behaviour well below the transition temperature and a very small influence on the magnetic transition. This conclusion should also apply to ferrimagnetic PBAs, in addition to ferromagnetic PBAs such as Co[Cr]$_{2/3}$. The magnetic MC model considers only first-nearest-neighbour interactions between P and R sites; therefore, changing the interaction from ferromagnetic to antiferromagnetic does not alter spin connectivity throughout the lattice. Consequently, vacancy arrangements should influence magnetic ordering similarly in both cases.

Although correlated vacancy order appears to have limited impact on magnetic order in PBAs, this is not universal among vacancy-ordered systems. In [Gua]Mn$_{1-x}$Fe$_{2x/3}\square_{x/3}$(HCOO)$_{3}$, long-range checkerboard vacancy ordering where $x>0.6$ leads to increased occupation of one magnetic sublattice over another \cite{BostromPerovskite2019}, driving a transition from antiferromagnetic to ferrimagnetic order \cite{Bulled2024}. Under checkerboard ordering, vacancies adopt an FCC arrangement analogous to the potentially vacant R sublattice in PBAs. However, vacancy distributions in antiferromagnetic PBAs would vary within this FCC sublattice without changing the relative occupation of the magnetic sublattices. Overall, vacancy ordering changes the relative occupation of the two magnetic sublattices in [Gua]Mn$_{1-x}$Fe$_{2x/3}\square_{x/3}$(HCOO)$_{3}$ leading to the onset of ferrimagnetic order, whereas in PBAs vacancies are rearranged within a single magnetic sublattice, leaving the ground-state magnetisation unchanged.

The absence of any effect of vacancy distributions on magnetisation in the MC model suggests that PBAs with identical composition but different vacancy arrangements should exhibit minimal differences in magnetic behaviour. Detailed characterisation of vacancy correlations may therefore be unnecessary when interpreting magnetism in PBAs, increasing confidence when comparing experimental magnetic data between PBAs sharing the same composition across literature. This situation contrasts with the strong dependence of mass transport on vacancy distributions \cite{Simonov2020}, which may arise from differences in the connectivity of pore and magnetic-site networks. Pore networks are sensitive to local vacancy distributions because vacancies only connect with their nearest neighbours and occupy only one-third of sites on an FCC sublattice, so a diversity of network types with different numbers of vacancy pairs per formula unit is possible, leading to varied mass transport properties \cite{Simonov2020}. In contrast, magnetic R sites can connect to their first and second nearest R site neighbours through P sites. Each R site is coordinated to six P sites, which can connect to a maximum of 18 further R sites. This high magnetic-site connectivity reduces the impact of vacancy-induced pathway disruption, as magnetic ordering can communicate via multiple pathways.

Whilst this simple MC model can describe PBAs demonstrating only one type of magnetic coupling, it does not accommodate more complex behaviour such as the influence of hydration on magnetic ordering behaviour observed experimentally in Ref.~\citenum{Ohkoshi2004}. The assumption that only one type of exchange interaction is present is incompatible with the model proposed by Ohksohi et al. that octahedral Co$^{\textrm{II}}$ sites in Co[Cr]$_{2/3}$ have ferromagnetic interactions with Cr$^{\textrm{III}}$ whilst tetrahedral Co$^{\textrm{II}}$ (postulated to be found when the PBA has decreased hydration) has antiferromagnetic interactions \cite{Ohkoshi2004}. The presence of these tetrahedral cobalt sites may only be possible when sites possess compatible Co$^{\textrm{II}}$--N coordination that can accommodate this change upon dehydration \cite{Ohkoshi2004}. As different vacancy distributions produce varying P-site coordination numbers and geometries, the proportion and distribution of potential tetrahedral sites could be influenced by vacancy correlations; this could change the fraction of antiferromagnetic interactions and therefore the magnetism. Incorporating hydration and multiple exchange interactions would therefore be required to encompass cases such as these and we leave explanation of these systems to future work.

\section*{Acknowledgements}

This research was supported financially by the European Research Council (Advanced Grant 788144), the UKRI through Frontier Research Grant EP/Z534031/1, the Royal Society through the Faraday Discovery Fellowships Fund, provided by DSIT, and the S.T.F.C. (studentship to E.A.H.). Experiments at the ISIS Neutron and Muon Source were supported by beamtime allocation RB2420345 from the Science and Technology Facilities Council. The raw diffraction data can be found at https://doi.org/10.5286/ISIS.E.RB2420345. We thank K.O.R. Juul for helping transport samples for preliminary measurements. 

  \bibliographystyle{elsarticle-num} 
  \bibliography{library.bib}

@article{wang_exploring_2026,
	title = {Exploring the {Role} of {Compositional} {Space} of {Disordered} {Rock} {Salt} {Na}-{Ion} {Battery} {Cathodes}},
	journal = {ACS Energy Lett.},
	author = {Wang, You and Watt, John and King, Madison and Liu, Rui and Nordlund, Dennis and Chen, Dongchang},
	year = {2026},
  
	pages = {3417--3426},
  

}

@article{dasilva_mean-field_2012,
	title = {A mean-field analysis of the exchange coupling ({J}) for noncubic {Prussian} blue analogue magnets},
	volume = {116},
	issn = {19327455},

	number = {46},
	journal = {J. Phys. Chem. C},
	publisher = {American Chemical Society},
	author = {DaSilva, Jack G. and McConnell, Amber C. and Fishman, Randy S. and Miller, Joel S.},
	year = {2012},
	pages = {24752--24756}
}

@article{Nielsen_2026,
	title = {Local structure of hydrated and dehydrated Prussian white cathode materials},
	volume = {14},
	number = {5},
	journal = {J. Mater. Chem. C},
	author = {Nielsen, I. and Eremenko, M. and Zhang, Y. and Tucker, M. G. and Brandt, W. R.},
	year = {2026},
	pages = {1912-1923}}

@article{Asselin2010,
   author = {P. Asselin and R. F.L. Evans and J. Barker and R. W. Chantrell and R. Yanes and O. Chubykalo-Fesenko and D. Hinzke and U. Nowak},

   issn = {10980121},
   issue = {5},
   journal = {Phys. Rev. B: Condens. Matter Mater. Phys.},
   month = {8},
   pages = {054415},
   title = {{Constrained Monte Carlo method and calculation of the temperature dependence of magnetic anisotropy}},
   volume = {82},
   year = {2010},
}

@article{Ohkoshi2023,
   author = {Shinichi Ohkoshi and Kosuke Nakagawa and Marie Yoshikiyo and Asuka Namai and Kenta Imoto and Yugo Nagane and Fangda Jia and Olaf Stefanczyk and Hiroko Tokoro and Junhao Wang and Takeshi Sugahara and Kouji Chiba and Kazuhiko Motodohi and Kazuo Isogai and Koki Nishioka and Takashi Momiki and Ryu Hatano},
 
   issn = {20411723},
   issue = {1},
   journal = {Nat. Commun.},
   month = {12},
   pages = {8466},
   pmid = {38151489},
   publisher = {Nature Research},
   title = {Giant adiabatic temperature change and its direct measurement of a barocaloric effect in a charge-transfer solid},
   volume = {14},
   year = {2023},
}

@article{Hu2022,
   author = {Yong Hu and Zipeng Guo and Yingjie Chen and Chi Zhou and Yuguang C. Li and Shenqiang Ren},

   issn = {20411723},
   issue = {1},
   journal = {Nat. Commun.},
   month = {12},
   pages = {7056},
   pmid = {36396649},
   publisher = {Nature Research},
   title = {Molecular magneto-ionic proton sensor in solid-state proton battery},
   volume = {13},
   year = {2022},
}

@article{Ono2017,
   author = {Kenta Ono and Manabu Ishizaki and Katsuhiko Kanaizuka and Takanari Togashi and Teppei Yamada and Hiroshi Kitagawa and Masato Kurihara},

   issn = {0044-8249},
   issue = {20},
   journal = {Angew. Chem. Int. Ed.},
   month = {5},
   pages = {5531-5535},
   publisher = {Wiley},
   title = {{Grain‐Boundary‐Free Super‐Proton Conduction of a Solution‐Processed Prussian‐Blue Nanoparticle Film}},
   volume = {56},
   year = {2017},
}

@article{Thallapally2010,
   author = {Praveen K. Thallapally and Radha Kishan Motkuri and Carlos A. Fernandez and B. Peter McGrail and Ghorishi S. Behrooz},

   issn = {00201669},
   issue = {11},
   journal = {Inorg. Chem.},
   month = {6},
   pages = {4909-4915},
   pmid = {20166714},
   title = {{Prussian Blue Analogues for CO$_2$ and SO$_2$ Capture and Separation Applications}},
   volume = {49},
   year = {2010},
}

@book{Young1993,
isbn = {0198555776},
language = {eng},
lccn = {LC 92-20900},
author = {Young, R. A. },
address = {Chester, England},
booktitle = {The Rietveld method},
series = {International Union of Crystallography monographs on crystallography ; 5},
title = {The Rietveld method },
year = {1993},
publisher = {International Union of Crystallograhy},
}

@article{Hurlbutt2018,
   author = {Kevin Hurlbutt and Samuel Wheeler and Isaac Capone and Mauro Pasta},

   issn = {25424351},
   issue = {10},
   journal = {Joule},
   month = {10},
   pages = {1950-1960},
   publisher = {Cell Press},
   title = {{Prussian Blue Analogs as Battery Materials}},
   volume = {2},
   year = {2018},
}

@article{Grandjean2016,
   author = {Fernande Grandjean and Louise Samain and Gary J. Long},

   issn = {14779234},
   issue = {45},
   journal = {Dalton Trans.},
   pages = {18018-18044},
   publisher = {Royal Society of Chemistry},
   title = {{Characterization and utilization of Prussian blue and its pigments}},
   volume = {45},
   year = {2016},
}

@article{Mugiraneza2022,
   author = {Sam Mugiraneza and Alannah M. Hallas},

   issn = {23993650},
   issue = {1},
   journal = {Commun. Phys.},
   month = {12},
   pages = {95},
   publisher = {Nature Research},
   title = {{Tutorial: a beginner’s guide to interpreting magnetic susceptibility data with the Curie-Weiss law}},
   volume = {5},
   year = {2022},
}

@article{Brown1969,
   author = {David Basset Brown and D. F. Shriver},
   issue = {1},
   journal = {Inorg. Chem.},
   pages = {37-42},
   title = {{Structures and Solid-State Reactions of Prussian Blue Analogs Containing Chromium, Manganese, Iron, and Cobalt}},
   volume = {8},

   year = {1969},
}

@article{Ludi1970,
   author = {Andreas Ludi and H.-U. Güdel and M Rüegg},
   journal = {Inorg. Chem.},
   pages = {2224-2227},
   title = {{The Structural Chemistry of Prussian Blue Analogs. A Single-Crystal Study of Manganese(II) Hexacyanocobaltate(III), Mn$_3$[Co(CN)$_6$]$_2$
.$x$H$_2$O}},
   volume = {9},

   year = {1970},
}

@article{Bhatt2013,
   author = {Pramod Bhatt and Nidhi Thakur and Mayuresh D. Mukadam and Sher Singh Meena and Seikh M. Yusuf},

   issn = {19327447},
   issue = {6},
   journal = {J. Phys. Chem. C},
   month = {2},
   pages = {2676-2687},
   title = {{Evidence for the existence of oxygen clustering and understanding of structural disorder in Prussian blue analogues molecular magnet M$_{1.5}$[Cr(CN)$_6$]$.z$H$_2$O (M = Fe and Co): Reverse Monte Carlo simulation and neutron diffraction study}},
   volume = {117},
   year = {2013},
}

@Article{Bostrom2024,
author ="Boström, Hanna L. B. and Cairns, Andrew B. and Chen, Muzi and Daisenberger, Dominik and Ridley, Christopher J. and Funnell, Nicholas P.",
title  ={{The pressure response of Jahn–Teller-distorted Prussian blue analogues}},
journal  ="Chem. Sci.",
year  ="2024",
volume  ="15",
issue  ="9",
pages  ="3155-3164",
publisher  ="The Royal Society of Chemistry"}

@article{Ramirez1991,
   author = {A. P. Ramirez},

   issn = {00218979},
   issue = {10},
   journal = {J. Appl. Phys.},
   pages = {5952-5955},
   title = {Thermodynamic measurements on geometrically frustrated magnets (invited)},
   volume = {70},
   year = {1991},
}

@article{chapon_wish_2011,
	title = {Wish: {The} {New} {Powder} and {Single} {Crystal} {Magnetic} {Diffractometer} on the {Second} {Target} {Station}},
	volume = {22},

	
	number = {2},
	journal = {Neutron News},
	publisher = {Taylor \& Francis},
	author = {Chapon, Laurent C. and Manuel, Pascal and Radaelli, Paolo G. and Benson, Chris and Perrott, Leigh and Ansell, Stuart and Rhodes, Nigel J. and Raspino, Davide and Duxbury, D. and Spill, E. and Norris, Julian},
	year = {2011},
	pages = {22--25},
}

@article{Herren1980,
   author = {F Herren and P Fischer and A Ludi and W Hälg},
   journal = {Inorg. Chem.},

   pages = {956-959},
   title = {{Neutron Diffraction Study of Prussian Blue, Fe$_4$[Fe(CN)$_6$]$_3$$.x$H$_2$O. Location of Water Molecules and Long-Range Magnetic Order}},
   volume = {19},

   year = {1980},
}

@article{Sato2003,
   author = {Yusuke Sato and S Ohkoshi and K Arai and Masanori Tozawa and Kazuhito Hashimoto},
   journal = {J. Am. Chem. Soc.},
   pages = {14590-14595},
   title = {{Solvatomagnetism-Induced Faraday Effect in a Cobalt Hexacyanochromate-Based Magnet}},
   volume = {125},
   year = {2003},

}

@article{Shante1971,
copyright = {Copyright Taylor \& Francis Group, LLC 1971},
issn = {0001-8732},
journal = {Adv. Phys.},
language = {eng},
number = {85},
author = {Shante, Vinod K.S. and Kirkpatrick, Scott},
title = {An introduction to percolation theory},
volume = {20},
year = {1971},
pages = {325-357},

publisher = {Taylor \& Francis Group},
}

@article{Ohkoshi2004,
   author = {S Ohkoshi and K Arai and Yusuke Sato and Kazuhito Hashimoto},
   journal = {Nat. Mater.},
   pages = {857-861},
   title = {Humidity-induced magnetization and magnetic pole inversion in a cyano-bridged metal assembly},
   volume = {3},
   year = {2004},

}

@incollection{Verdaguer2004,
author = {Verdaguer, Michel and Girolami, Gregory S.},
publisher = {John Wiley \& Sons, Ltd},
isbn = {9783527604388},
title = {{Magnetic Prussian Blue Analogs}},
booktitle = {Magnetism: Molecules to Materials V},
chapter = {9},
pages = {283-346},



year = {2004},
editor = {J.S. Miller and M. Drillon}
}

@article{Kaye2005,
   author = {Steven S. Kaye and Jeffrey R. Long},

   issn = {00027863},
   issue = {18},
   journal = {J. Am. Chem. Soc.},
   month = {5},
   pages = {6506-6507},
   title = {Hydrogen storage in the dehydrated Prussian blue analogues {M}$_3$[{Co}({CN})$_6$]$_2$ ({M = Mn, Fe, Co, Ni, Cu, Zn})},
   volume = {127},
   year = {2005},
}

@article{Kim2009,
   author = {J. E. Kim and K. Kato and M. Takata and T. Shibata and Y. Moritomo},

   issn = {10980121},
   issue = {13},
   journal = {Phys. Rev. B: Condens. Matter Mater. Phys.},
   month = {4},
   title = {{Guest-host interaction of K$_{0.34}$Co[Fe(CN)$_6$]$_{0.75}.z$H$_2$O as investigated by a charge-density analysis}},
   volume = {79},
   year = {2009},
   pages = {132105}
}

@article{Reguera2010,
   author = {E Reguera and J Rodríguez-Hernández and C Tellez and M Centeno},
   journal = {Z. Phys. Chem.},
   pages = {807-826},
   title = {{On the Low Stability of Molecular Magnets Based on Transition Metal Hexacyanochromates (III)}},
   volume = {224},
   year = {2010},
}

@article{Ohkoshi2010,
   author = {S Ohkoshi and Kosuke Nakagawa and Keisuke Tomono and Kenta Imoto and Yoshihide Tsunobuchi and Hiroko Tokoro},
   journal = {J. Am. Chem. Soc.},

   pages = {6620-6621},
   title = {{High Proton Conductivity in Prussian Blue Analogues and the Interference Effect by Magnetic Ordering}},
   volume = {132},
   year = {2010},
}

@article{Pajerowski2015,
   author = {D. M. Pajerowski and S. E. Conklin and J. Leão and L. W. Harriger and D. Phelan},

   issn = {1550235X},
   issue = {9},
   journal = {Phys. Rev. B: Condens. Matter Mater. Phys.},
   month = {3},
  
   pages = {094104},
   publisher = {American Physical Society},
   title = {{High-pressure neutron scattering of the magnetoelastic Ni-Cr Prussian blue analog}},
   volume = {91},
   year = {2015},
}

@article{Marquez2018,
   author = {Carlos Marquez and Francisco G. Cirujano and Cédric Van Goethem and Ivo Vankelecom and Dirk {{De Vos}} and Trees {{De Baerdemaeker}}},

   issn = {20444761},
   issue = {8},
   journal = {Catal. Sci. Technol.},
   pages = {2061-2065},
   publisher = {Royal Society of Chemistry},
   title = {{Tunable Prussian blue analogues for the selective synthesis of propargylamines through A$^3$ coupling}},
   volume = {8},
   year = {2018},
}

@article{JTBostrom2019,
   author = {Hanna L.B. Boström and Ines E. Collings and Andrew B. Cairns and Carl P. Romao and Andrew L. Goodwin},

   issn = {14779234},
   issue = {5},
   journal = {Dalton Trans.},
   pages = {1647-1655},
   pmid = {30548036},
   publisher = {Royal Society of Chemistry},
   title = {{High-pressure behaviour of Prussian blue analogues: Interplay of hydration, Jahn-Teller distortions and vacancies}},
   volume = {48},
   year = {2019},
}

@article{Simonov2020,
   author = {Arkadiy Simonov and Trees {{De Baerdemaeker}} and H. L. B Boström and M. L {{Ríos Gómez}} and H. J. Gray and D. Chernyshov and A. Bosak and H-B Bürgi and A. L. Goodwin},
   journal = {Nature},
   pages = {256-260},
   title = {{Hidden diversity of vacancy networks in Prussian blue analogues}},
   volume = {578},
   year = {2020},

}

@article{Zampardi2020,
   author = {Giorgia Zampardi and Fabio La Mantia},
 
   issn = {24519111},
   journal = {Curr. Opin. Electrochem.},
   month = {6},
   pages = {84-92},
   publisher = {Elsevier B.V.},
   title = {{Prussian blue analogues as aqueous Zn-ion batteries electrodes: Current challenges and future perspectives}},
   volume = {21},
   year = {2020},
}

@article{Cattermull2021,
   author = {John Cattermull and Mauro Pasta and Andrew L. Goodwin},
   journal = {Mater. Horiz.},
   pages = {3178-3186},
   title = {{Structural complexity in Prussian blue analogues}},
   volume = {8},
   year = {2021},

}

@book{Blundell2001,
Author = {Blundell, S.},
ISBN = {9780198505921},
Publisher = {OUP Oxford},
Series = {Oxford Master Series in Condensed Matter Physics},
Title = {Magnetism in Condensed Matter.},

Year = {2001},
}

@article{paddison_ewald,
author = {Joseph A. M. Paddison  and Henrik Jacobsen  and Oleg A. Petrenko  and Maria Teresa Fernández-Díaz  and Pascale P. Deen  and Andrew L. Goodwin },
title = {Hidden order in spin-liquid {Gd$_3$Ga$_5$O$_{12}$}},
journal = {Science},
volume = {350},
number = {6257},
pages = {179-181},
year = {2015}}

@article{Bostrom2021,
   author = {Hanna L.B. Boström and Ines E. Collings and Dominik Daisenberger and Christopher J. Ridley and Nicholas P. Funnell and Andrew B. Cairns},

   issn = {15205126},
   issue = {9},
   journal = {J. Am. Chem. Soc.},
   month = {3},
   pages = {3544-3554},
   pmid = {33629831},
   publisher = {American Chemical Society},
   title = {{Probing the Influence of Defects, Hydration, and Composition on Prussian Blue Analogues with Pressure}},
   volume = {143},
   year = {2021},
}

@article{Yin2022,
   author = {Yan Yin and Mengyuan Li and Xin Liu and Na Xie and Chenyang Zheng and Guokun Liu and Shuoyao Yin and Xintian Li and Yabiao Pei and Junfeng Zhang and Yanzhou Qin and Michael D. Guiver},

   issn = {03787753},
   journal = {J. Power Sources},
   month = {7},
   pages = {231542},
   publisher = {Elsevier B.V.},
   title = {{Cost-effective Prussian blue analogue composite proton exchange membranes for low humidity fuel cell operation}},
   volume = {537},
   year = {2022},
}

@incollection{Binney1992,
language = {eng},
publisher = {Oxford University Press, Incorporated},
title = {Introduction},
chapter = {1},
year = {1992},
author = {Binney, J. J and Dowrick, N. J and Fisher, A. J and Newman, M. E. J},
address = {United Kingdom},
booktitle = {The Theory of Critical Phenomena},
isbn = {9780198513933},
}

@book{TOPAS,
    author = {Alan A Coelho},
    publisher = {Coelho Software: Brisbane},
    title = {{TOPAS-Academic, version 6 (computer software)}},
    volume = {},
}

@book{ISODISTORT1,
    author = {H. T. Stokes and D. M. Hatch and B. J. Campbell},
    publisher = {iso.byu.edu},
    title = {{ISODISTORT, ISOTROPY Software Suite}},
    volume = {},
    year = {},
    
 
}

@article{ISODISTORT2,
   author = {B. J. Campbell and H. T. Stokes and D. E. Tanner and D. M. Hatch},
   journal = {J. Appl. Cryst.},
   pages = {607-614},
   title = {{ISODISPLACE: An Internet Tool for Exploring Structural Distortions.}},
   volume = {39},
   year = {2006},
}

@article{Gawron1991,
   author = {T R Gawron},
   journal = {Acta Phys. Pol., A},
   pages = {461-464},
   title = {{Site percolation thresholds of FCC lattice}},
   volume = {80},
   year = {1991}

}

@article{Mihalik2010,
   author = {M. Mihalik and V. Kavečanský and S. Mat'Aš and M. Zentková and O. Prokhenko and G. André},

   issn = {17426596},
   issue = {SECTION 2},
   journal = {J. Phys.: Conf. Ser.},
   pages = {022035},
   publisher = {Institute of Physics Publishing},
   title = {{Magnetic properties and neutron diffraction study of (Ni$_x$Mn$_{1-x}$)$_3$[Cr(CN)$_6$]$_2$ molecule-based magnets}},
   volume = {200},
   year = {2010},
}

@article{Ferlay1995,
   author = {S. Ferlay and T. Mallah and R. Ouahes and P. Veillet and M. Verdaguer},
   journal = {Nature},
   pages = {701-703},
   title = {{A room-temperature
organometallic magnet
based on Prussian blue}},
   volume = {378},
   year = {1995},
}

@article{Zentkova2006,
   author = {M. Zentková and M. Mihalik and J. Kováč and A. Zentko and Z. Mitróová and M. Lukáčová and V. Kavečanský and L. F. Kiss},
   journal = {Phys. Stat. Sol. B},
   pages = {272-276},
   title = {{Magnetic properties of TM$_3$[Cr(CN)$_6$]$_2 . n$H$_2$O}},
   volume = {243},
   year = {2006},
 
}

@article{bhatt_multifunctional_2025,
	title = {Multifunctional {Properties} and {Potential} {Applications} of {Prussian} {Blue} {Analogue} {Magnets}},
	volume = {37},

	journal = {Chem. Mater.},
	author = {Bhatt, Pramod and Kumar, Amit and Yusuf, S. M.},
	year = {2025},
	pages = {551--576}

}

@article{Pajerowski2012,
   author = {Daniel M. Pajerowski and V. Ovidiu Garlea and Elisabeth S. Knowles and Matthew J. Andrus and Matthieu F. Dumont and Yitzi M. Calm and Stephen E. Nagler and Xin Tong and Daniel R. Talham and Mark W. Meisel},

   issn = {10980121},
   issue = {5},
   journal = {Phys. Rev. B: Condens. Matter Mater. Phys.},
   month = {8},
   pages = {054431},
   title = {{Magnetic neutron scattering of thermally quenched K-Co-Fe Prussian blue analog photomagnet}},
   volume = {86},
   year = {2012},
}

@article{Kumar2008,
   author = {Amit Kumar and S. M. Yusuf and L. Keller and J. V. Yakhmi},
   journal = {Phys. Rev. Lett.},
   pages = {207206},
   title = {{Microscopic Understanding of Negative Magnetization in Cu, Mn, and Fe Based Prussian Blue Analogues}},
   volume = {101},
   year = {2008},

}

@article{Nakotte2016,
   author = {Heinz Nakotte and Manjita Shrestha and Sourav Adak and Michael Boergert and Vivien S. Zapf and Neil Harrison and Graham King and Luke L. Daemen},

   issn = {24682179},
   issue = {2},
   journal = {J. Sci.: Adv. Mater. Devices},
   month = {6},
   pages = {113-120},
   publisher = {Elsevier B.V.},
   title = {{Magnetic properties of some transition-metal Prussian Blue Analogs with composition M$_3$[M'(C,N)$_6$]$_2·x$H$_2$O}},
   volume = {1},
   year = {2016},
}

@article{Broadbent1957, title={{Percolation processes: I. Crystals and mazes}}, volume={53},  number={3}, journal={Math. Proc. Cambridge Philos. Soc.}, author={Broadbent, S. R. and Hammersley, J. M.}, year={1957}, pages={629–641}}

@article{Bulled2024,
author = {Bulled, Johnathan M. and Willis, Alexandra and Faure Beaulieu, Zoé and Cassidy, Simon J. and Bruckmoser, Jonas and Boström, Hanna L. B. and Goodwin, Andrew L.},
title = {{Percolation-Induced Ferrimagnetism from Vacancy Order in [Gua]Mn$_{1–x}$Fe$_{2x/3}$(HCOO)$_3$ Hybrid Perovskites}},
journal = {J. Am. Chem. Soc.},
year = {2024},
volume={146},
pages = {13714–13718}

}

@article{BostromPerovskite2019,
   author = {Hanna L.B. Boström and Jonas Bruckmoser and Andrew L. Goodwin},

   issn = {15205126},
   issue = {45},
   journal = {J. Am. Chem. Soc.},
   month = {11},
   pages = {17978-17982},
   pmid = {31663731},
   publisher = {American Chemical Society},
   title = {{Ordered B-Site Vacancies in an ABX$_3$ Formate Perovskite}},
   volume = {141},
   year = {2019},
}

@article{Metropolis1949,
   author = {Nicholas Metropolis and S Ulam},
   issue = {247},
   journal = {J. Am. Stat. Soc},
   pages = {335-341},
   title = {{The Monte Carlo Method}},
   volume = {44},
   year = {1949},

}

@article{Davidson1928,
   author = {David Davidson and Lars A Welo},
   journal = {J. Phys. Chem.},
   pages = {1191-1196},
   publisher = {UTC},
   title = {{The nature of Prussian blue}},
   volume = {32},

   year = {1928},
}
\end{document}